# Suppressing the fragmentation of fragile molecules in helium nanodroplets by co-embedding with water: Possible role of the electric dipole moment

Yanfei Ren and Vitaly V. Kresin

*Department of Physics and Astronomy, University of Southern California,*

*Los Angeles, California 90089-0484*

**ABSTRACT**

When fragile molecules such as glycine, polyglicine, alkanes, and alkanethiols are embedded in liquid helium nanodroplets, electron-impact ionization of the beam leads to fragmentation which is as extensive as that of isolated gas-phase molecules. However, it turns out that if a few molecules of water are co-embedded with the peptide and alkane chains, their fragmentation is drastically reduced or completely eliminated. On the other hand, the fragmentation of alkanethiols remains unaffected. On the basis of these observations, it is proposed that the fragmentation "buffering" effect may correlate with the magnitude of the impurity's electric dipole moment, which steers the migration of the ionizing $He^+$ hole in the droplet.

The ability of nanoscale $^4$He droplets flying in a molecular beam to pick up diverse foreign atoms and molecules, embed and carry them along, cool them to a sub-Kelvin temperature, and make them available for precise spectroscopic measurements has given rise to a burst of research exploring this "ultimate spectroscopic matrix" [1].

In the beginning of helium nanodroplet studies, it was hoped that embedding fragile species in the middle of these "cryopillows" would facilitate fragmentation-free mass spectrometry. However, this turned out not to be the case. When doped droplets are probed by electron bombardment, the ionization of impurities proceeds primarily via charge transfer to the embedded species from a positive He$^+$ hole migrating through the droplet (see, e.g., [2-4]). The accompanying energy release is substantial, and often results in extensive fragmentation of the impurity (see, e.g., the overview in Refs. [5,6]). For example, it has been found that glycine (the lightest amino acid molecule), and tryptophan (the heaviest) survive ionization inside He$_n$ only marginally better than in the gas phase [7-9].

In the following, we demonstrate that the presence of just a few water molecules embedded in a helium droplet next to short polyglycine chains dramatically suppresses their ionization-induced fragmentation. A companion measurement was performed on non-polar and polar hydrocarbon chains (alkanes and alkanethiols), allowing us to propose a correlation between this "buffering" effect and the magnitudes of the electric dipole moments of co-embedded impurities.

Mass spectra shown below were acquired using a supersonic beam of $^4$He$_n$ droplets produced using a 5 μm nozzle at $T$=14 K and $P$=40 bar. These conditions match earlier experiments[7,8] and correspond to $\langle n \rangle \approx 10^4$ [10]. In the pick-up chamber the beam passed though a chopper followed by two pick-up cells. The first was fed by a stable pressure of H$_2$O or D$_2$O vapor through a narrow tubing, and the second contained glycine, alkanes, or alkanethiols. When filled with glycine oligomer powder, this pick-up cell was heated to 160-200 °C; no serious thermal decomposition was indicated by the mass spectra. The gas C$_3$H$_8$ and the liquid vapors C$_3$H$_8$S, C$_6$H$_{14}$, and C$_6$H$_{14}$S were admitted via a length of narrow tubing through a momentarily opened valve, and the resulting charge was sufficiently stable over the approximately three-minute duration of the subsequent mass scan. According to the mass spectra, the droplets picked up on average 3 water molecules, while vapor pressures in the second cell were kept low enough to pick up only single molecules. As will be described below, control runs also were conducted with different source conditions and with reversed pick-up order. In the final chamber the doped droplets were detected by a quadrupole mass spectrometer synchronized with the chopper. The mass distributions remained unaltered for electron energies scanned from 28 to 60 eV.

Fig. 1 shows that ionization-induced fragmentation of diglycine inside $He_n$ is extensive. In fact, the products are essentially the same as for isolated gas phase molecules [11] with only minor changes in the relative branching ratios. There is no intact analyte molecule present in the mass spectrum. The same outcome holds for tri- and tetraglycine [12] and for individual glycine molecules, as mentioned above.

Upon co-embedding a few water molecules, the mass spectra change drastically. Fig. 1 shows that in the case of diglycine all the previous fragmentation peaks disappear, and only the pure diglycine and its complexes with water are present. Replacing regular water with $D_2O$ confirmed that protonation comes only from hydroxyl loss by the water molecules and not from diglycine. This reveals that fragmentation of the latter is completely buffered by the proximity of water, as was also observed for single glycine molecules in our previous experiment [8].

For triglycine, Fig. 2, the buffering effect of water is also very robust. Only one fragment remains (at 114 amu, assigned to the dimer $(NHCH_2CO)_2^+$ which is one of the principal fragmentation products for single molecules both in the gas phase and in nanodroplets [11,12]), and a strong intact triglycine ion is observed at 189 amu. However, this time there are no additional hydrated peaks.

Whereas mono- through triglycine fragmentation is suppressed by the presence of water in a qualitatively similar way, Fig. 2 shows that tetraglycine fragmentation is no longer buffered by the picked-up quantity of water. There is no intact molecular ion, and the major fragmentation peaks [11,12] have reappeared.

These outcomes were found to be insensitive to changes in the pick-up sequence (i.e., whether droplets pick up water or glycine molecules first), the electron bombardment energy, and the mean $^4He_n$ droplet size (from $\langle n \rangle \approx 7 \times 10^3$ to $\sim 10^6$). The latter fact implies that the changes taking place as the peptide size increases are not related to the accompanying shrinkage of the host droplets as they evaporatively cool off the increasing number of thermally excited vibrational degrees of freedom of their guests.

What could underlie the dramatic reduction in the fragmentation of the glycine molecule and its smaller oligomers occasioned by the presence of a few water molecules in the droplet? This could be a solvation effect, whereby water molecules surround and protect the fragile species (as has been seen, for example, for nucleotides solvated in bare water clusters [13]). Consider, however, that the mass spectra remain unchanged if water is picked up first, in which case it presumably forms clusters [14,15] prior to the arrival of glycine. If solvation is dominantly responsible for the fragmentation buffering effect, it would have to occur in this situation as well, which would require the large glycine structures to jostle into the preformed water clusters in the cryogenic droplet. This possibility cannot be excluded (cf. the apparent ability of picked-up water molecules to insert themselves into preformed cyclic rings [14,16]), but it is not obvious and does not easily correlate with the alkane observations described below.

On the other hand, if solvation is not the dominant effect and the peptides end up parked near the water complexes, the mass spectra imply that it is the latter that are the preferred targets for the migrating $He^+$. In this case, the relative magnitude of the targets' electric dipole moments may account for the difference. Indeed, is has been shown that steering of charge migration in a nanodroplet by the multipole electrostatic potential of an embedded impurity is a quantitatively correct picture [17]. The water molecule has a dipole moment of $p=1.9$ D, and small water clusters in their ground states have $p\approx 1.1-3$ D [18,19]. The lowest-energy neutral conformation of glycine, on the other hand, has $p=0.7-0.9$ D [20]. (We were unable to find reliable experimental values for neutral polyglycines; their dipole moments may be expected to increase with size, but relatively slowly [21].) Thus ionization steering by the dipole field may enable the observed restructuring of the ionization mass spectra for mono-, di- and triglycine molecules. The recurrence of strong fragmentation for tetraglycine is consistent both with its large size and with its larger dipole moment acting to return it to favored-target status for $He^+$.

In order to examine further whether dipole moments are indeed relevant to the ability of water to shield impurities from dissociation, it is useful to perform comparative measurements on molecules that are similar in structure but different in the strength of their electric dipoles. A convenient family of such molecules is offered by the linear $n$-alkanes $CH_3$-$(CH_2)_n$-$CH_3$ and alkanethiols $CH_3$-$(CH_2)_n$-$SH$. While alkanes have symmetric molecular conformations which lead to near-zero electric dipole moments, the asymmetric alkanethiols possess rather high moments [22]. We investigated two pairs of different lengths: propane ($p=0.084$ D) and 1-propanethiol (1.6 D), and hexane (0.09 D) and 1-hexanethiol (whose dipole moment should be close to that of propanethiol [23]).

As with polyglycine molecules, all of these systems were found to suffer the same degree of fragmentation when ionized within helium droplets as they do in the gas phase [12,24]. However, when co-embedded with (heavy) water, the nonpolar and polar chains exhibited very different behavior. Figs. 3 and 4 show that the presence of water strongly suppresses the ionization-induced fragmentation of propane and hexane, while having no significant effect at all on propanethiol and hexanethiol fragmentation. As with polyglycines, the effect did not vary with the electron energy or droplet size.

While shielding of alkane molecules by solvation again cannot be excluded by the data, it is noteworthy that these molecules are hydrophobic. In fact, if solvation made the main contribution one might expect the water molecules to be stronger attracted to the polar group of the alkanethiols and protect them more effectively, which is the opposite of what is observed. Therefore it appears that the fragmentation buffering effect is indeed strongly correlated with the magnitude of the electric dipole moment of the impurity, as suggested above.

In summary, it has been found that ionization-induced fragmentation of a fragile molecule carried by a liquid helium nanodroplet can be drastically reduced by co-embedding a few water molecules next to it. This new effect was observed for glycine molecules and peptides up to triglycine, and for alkane chains. On the other hand, no fragmentation protection transpires for the structurally similar but strongly dipolar alkanethiols. The results are consistent with the supposition that the extent of impurity fragmentation correlates with the magnitude of its electric dipole moment which can guide the migration of the ionizing charge.

The conspicuous nature of the fragmentation buffering effect would make it interesting to explore the nature of the formed complexes and their ionization dynamics further, both theoretically and spectroscopically. The unique ability afforded by the helium nanodroplet pick-up technique to generate atomic and molecular complexes of arbitrary composition suggests that variants of this effect produced with other combinations of polar and non-polar molecules may be of value for spectroscopy and mass spectrometry.

We would like to thank Prof. D. T. Moore for useful suggestions. This work was supported by NSF under grant No. PHY-0245102.

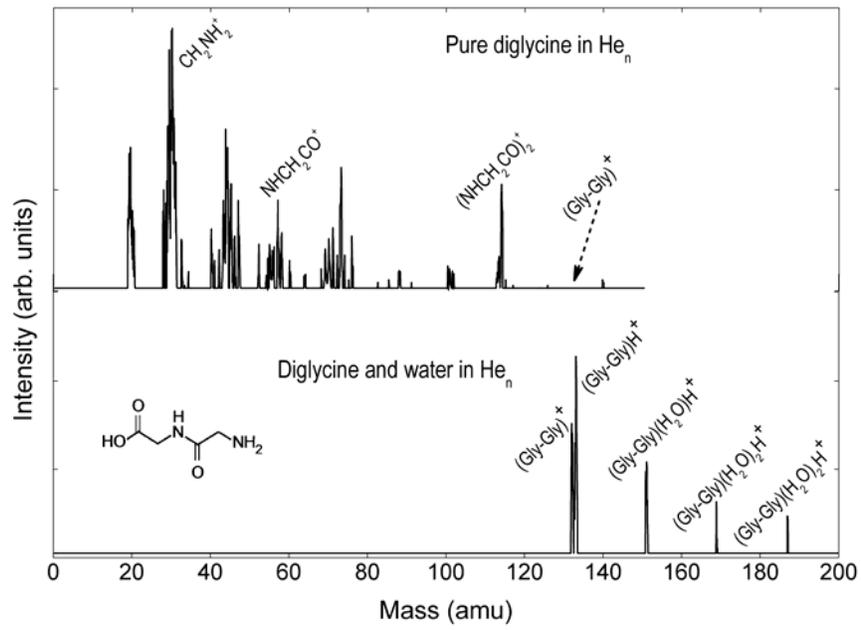

Figure 1

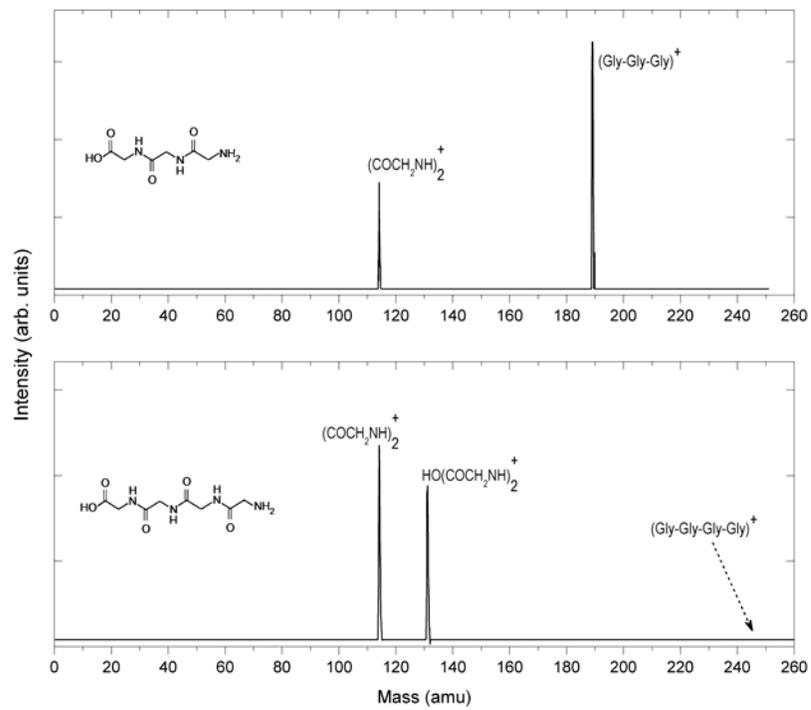

Figure 2

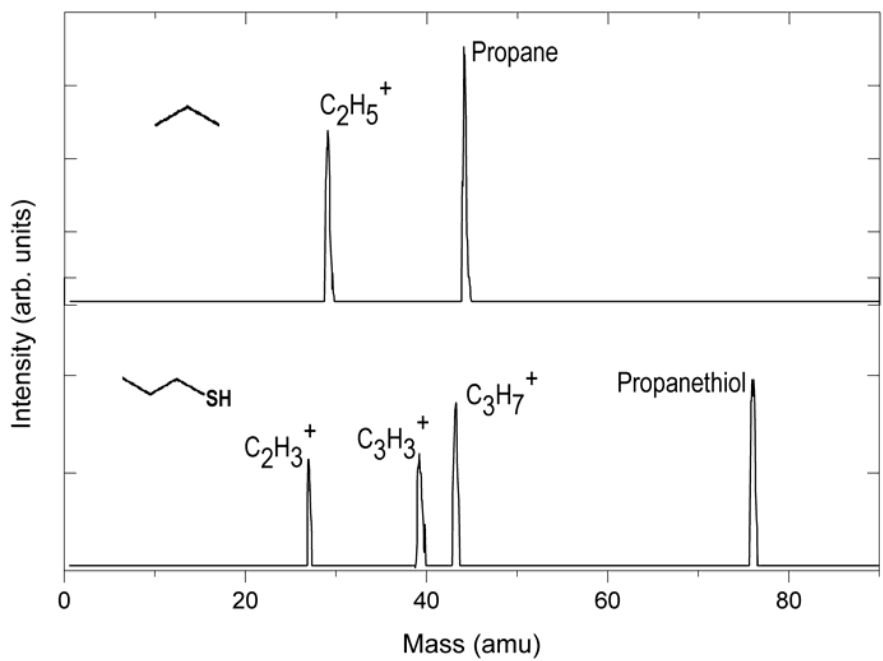

Figure 3

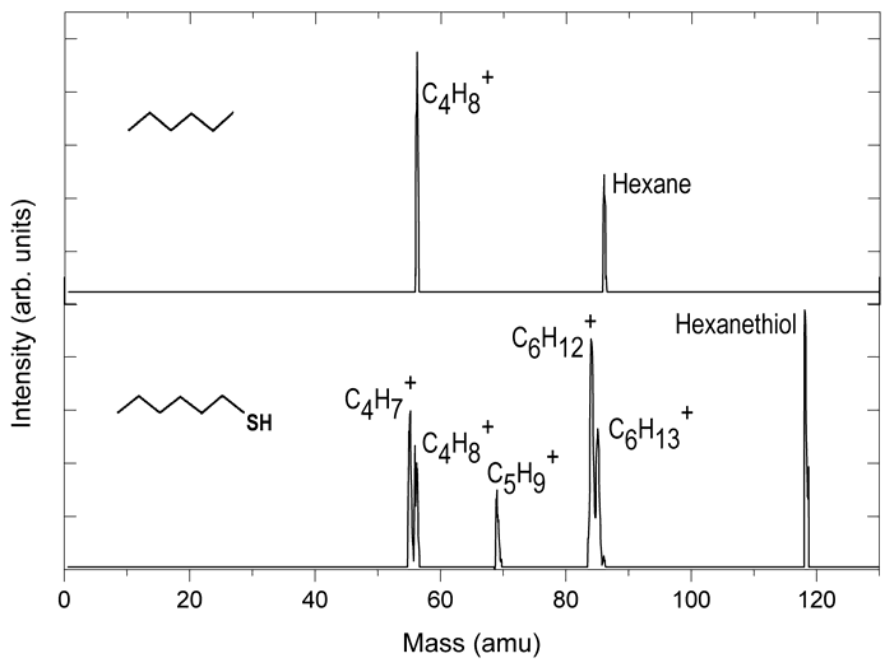

Figure 4

# Figure captions

**Figure 1.** *Top*: Electron impact ionization mass spectrum of He droplets doped with diglycine only. There is extensive fragmentation, with no observable signal at the parent mass. The spectrum closely resembles that of gas-phase diglycine, differing only in the magnitude of some branching ratios. *Bottom*: Mass spectrum obtained for He droplets picking up $(H_2O)_n$ followed by diglycine. There is no fragmentation, and only intact and hydrated diglycine peaks are seen. The mass spectra in this and all the other figures were acquired with an electron energy of 50-60 eV, and for clarity all the peaks unrelated to diglycine and its fragments (such as individual helium and water ions) were removed from the plots.

**Figure 2.** *Top*: Pattern resulting from He droplets picking up water followed by triglycine. The parent ion is present, together with a relatively small fragment peak; no parent+$(H_2O)_n$ complexes are observed. *Bottom*: The same for tetraglycine. This time there are two strong fragment peaks and no sign of the parent ion. In both cases, if no water is present in the droplet, fragmentation of the polyglycine molecules is found to be much more extensive and similar to that of free gas-phase molecules (cf. Fig. 1).

**Figure 3.** *Top*: Pattern resulting from He droplets doped with water followed by propane. In the absence of water in the droplet, $C_3H_8$ fragments extensively, but in the present case only one fragment peak remains. *Bottom*: Embedding propanethiol instead of propane together with water. Fragmentation is no longer suppressed, and the mass spectrum resembles that without water and that of a gas-phase molecule.

**Figure 4.** The patterns for hexane and hexanethiol co-embedded with water are similar to the situation in Fig. 3.

# References

[1] K. K. Lehmann and G. Scoles, Science **279**, 2065 (1998).

[2] A. Scheidemann, B. Schilling, and J. P. Toennies, J. Phys. Chem. **97,** 2128 (1993).

[3] B. E. Callicoatt, D.D. Mar, V.A. Apkarian, and K.C. Janda, J. Chem. Phys. **105,** 7872 (1996).

[4] M. Fárník and J. P. Toennies, J. Chem. Phys. **122,** 014307 (2005).

[5] S. Yang, S. M.Brereton, M. D.Wheeler, and A. M. Ellis, Phys. Chem. Chem. Phys. **7,** 4082 (2005).

[6] S. Yang, S. M. Brereton, and A. M. Ellis, Intern. J. Mass Spectrom. **253**, 79 (2006).

[7] F. Huisken, O. Werhahn, A. Yu. Ivanov, and S. A Krasnokutski, J. Chem. Phys. **111,** 2978 (1999).

[8] Y. Ren, R. Moro, and V. V. Kresin, Eur. Phys. J. D **43**, 109 (2007).

[9] A. Lindinger, J. P. Toennies, and A. F. Vilesov, J. Chem. Phys. **110,** 1429 (1999).

[10] J. P. Toennies and A. F. Vilesov, Angew. Chem. Int. Ed. **43**, 2622 (2004).

[11] NIST Chemistry WebBook, ed. by P. Linstrom and W. G. Mallard (National Institute of Standards and Technology, Gaithersburg, 2003), http://webbook.nist.gov/chemistry.

[12] Y. Ren, Ph. D. thesis, University of Southern California (2007).

[13] B. Liu, S. Brøndsted Nielsen, P. Hvelplund, H. Zettergren, H. Cederquist, B. Manil and B. A. Huber, Phys. Rev. Lett. **97**, 133401 (2006).

[14] K. Nauta and R. E. Miller, Science **287**, 293 (2000).

[15] M. N. Slipchenko, K. E. Kuyanov, B. G. Sartakov, and A. F. Vilesov, J. Chem. Phys. **124**, 241101 (2006).

[16] C. J. Burnham, S. S. Xantheas, M. A. Miller, B. E. Applegate, and R. E. Miller, J. Chem. Phys. **117,** 1109 (2002).

[17]. W. K. Lewis, C. M. Lindsay, R. J. Bemish, and R.E. Miller, J. Am. Chem. Soc. **127,** 7235 (2005).

[18] M. Yang, P. Senet, and C. Van Alsenoy, Int. J. Quantum Chem. **101,** 535 (2005).

[19] J. K. Gregory, D. C. Clary, K. Liu, M. G. Brown, and R. J. Saykally, Science **275, 814** (1997).

[20] F. J. Lovas Y. Kawashima, J.-U. Grabow, R. D. Suenram, G. T. Fraser, and E. Hirota, Astrophys. J. **455**, L201 (1995).

[21] P. Chaudhuri and S. Canuto, J. Mol. Struc. -THEOCHEM **760**, 15 (2006).

[22] *CRC Handbook of Chemistry and Physics*, 82nd ed., ed. by D. R. Lide (CRC Press, Boca Raton, 2001).

[23] S. Mathias and E. de Carvalho Filho, J. Phys. Chem. **62**, 1427 (1958).

[24] An analogous degree of fragmentation has been seen for dodecanethiol by J. D. Close, K. G. H. Baldwin, K. Hoffmann, and N Quaas, Appl. Phys. B **70**, 651 (2000).